\pdfoutput=1
\documentclass{ws-rv961x669}
\usepackage{ws-rv-van}     
\usepackage{ws-rv-thm}     
\makeindex

\begin{document}

\chapter[]{Fluctuation-dominated Phase Ordering}

\author[]{Mustansir Barma}

\address{TIFR Centre for Interdisciplinary Sciences,\\
Tata Institute of Fundamental Research,\\
Hyderabad 500046, India\\ 
barma23@gmail.com}

\begin{abstract}
Fluctuation-dominated phase ordering refers to a steady state in which the magnitude of long-range order varies strongly owing to fluctuations, and to the associated coarsening phenomena during the approach to steady state. Strong fluctuations can lead to a number of interesting phenomena, including a cusp singularity in the scaled correlation function, implying the breakdown of the Porod Law. First identified in a nonequilibrium system of passively sliding particles on a fluctuating surface, fluctuation-dominated order also occurs in several other systems, including an equilibrium Ising model with long-range interactions. This article discusses these systems,  and others where clustering effects are stronger.
\end{abstract}


\body 


\section{Meeting and working with Michael Fisher}\label{ra_sec1}
As a graduate student at Stony Brook in the early 1970s, I first heard about Michael Fisher from Helen 
Au Yang, who was then a senior student working with Barry McCoy. She recommended strongly that I read Michael’s 1967 Reports on Progress in Physics review of critical phenomena \cite{fisher67}. This was my
 introduction to the subject, and I was captivated by Michael's description of singularities, scaling and universality,   and the diversity of phenomena they describe.

My first encounter with Michael in person happened a few years later. I attended  a summer school on
 critical phenomena and the renormalization group organized by Donald Betts in Banff, where Michael
 was one of the lecturers. Younger participants were asked to give short talks on their work, and I
 talked about the isostructural first order phase transition in samarium sulphide, using Ising variables to 
 represent the two possible valencies of the samarium ions. A prominent experimentalist criticized me
 strongly, saying the model was not realistic enough. But before I could say anything in reply, Michael
 spoke out strongly about the importance of simple models in clarifying complex phenomena, and 
saved me from having to respond! 

Then in the early 1980s, I had an opportunity to work with Michael when I spent a couple of years at
 Cornell as a research associate, on leave from the Tata Institute of Fundamental Research (TIFR) in
 Mumbai (then Bombay). We worked on the application of partial differential approximants to study
  singular corrections to scaling in two dimensional Ising-like systems, and the Gaussian to Ising
   crossover which showed some unusual features. It was a rewarding experience, and I learned a lot. Michael took every bit of science very seriously and was able to get the
 best out of his collaborators, no matter what their backgrounds or skills. While preparing a
  manuscript, he was meticulous. Every word had to be just right, and the figures I prepared would
  invariably be re-drawn by him before being sent to the drafting section. He was extremely caring about 
  the welfare of his students and postdocs.  During summers, I was asked to interact with students who
   were yet to start on doctoral research. I wrote a paper with one of them, Steve White, which we
    requested Michael to read. It came back covered in red, with remarks all over and arrows to mark
     transpositions, along with the admonition “Don’t put the cart before the horse!”.  

I learned a lot from him, physics-wise and otherwise, both abstract and practical. He was extremely
 organized, and instructed me on the virtues of coloured separators in files (hard-bound, not
  soft). A discussion I remember well had to do with loyalty. Michael remarked that in addition to loyalty
  to the family and to the country, it is very important to be loyal to one’s profession. In this connection, 
 one should never refuse invitations to go and give lectures, including at colleges and schools. Over
 the years, that bit of advice has stayed with me and I have passed it on to my students, along with many other points I learned from him.

\begin{figure}[ht]
\centerline{\includegraphics[width=11cm]{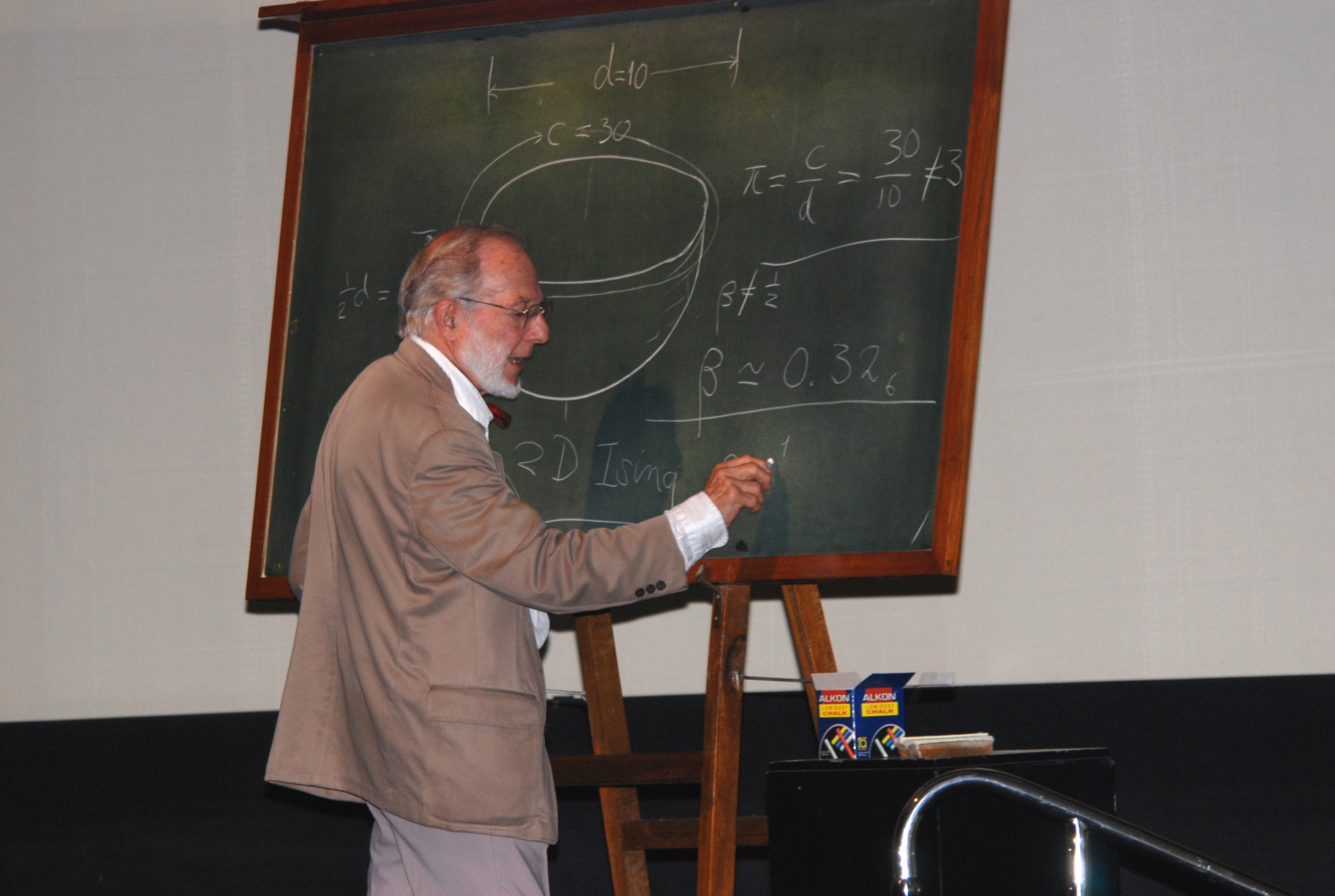}}
\caption{Michael Fisher delivering the Homi Bhabha lecture at the Tata Institute of Fundamental Research, Mumbai in 2007. Source: Tata Institute of Fundamental Research. This photograph first appeared in Current Science, {\bf123}, 817 (2022).
}
\label{fig8}
\end{figure}

In 2007 he visited India and delivered a special commemorative lecture, at the Tata Institute.  The 
subject was phase transitions (Fig. 1). The lecture was preceded by a good deal of preparation by the 
auditorium staff, who had been asked to ensure there was a table with a minimum specified length, and
a pointer which was long enough. Michael later told me that he found the arrangements at TIFR
 amongst the best he had seen anywhere. But the arrangements at the hotel he stayed at with Sorrel 
 and her sister Jacinta, were not quite to his satisfaction. As was his wont, Michael moved the furniture
  in the hotel room around until he felt it was right. This sort of rearrangement was not unexpected, as it
   was an unusual but endearing habit that he was well known for \cite{mermin}.

\section{Introduction}
 
 Michael’s work encompassed the subject of phase transitions and critical phenomena in a wide variety
 of systems, ranging from multicritical points where several critical loci meet, to multiphase points from
 which emanate an infinite number of phases. As emphasized by him, criticality defines a new state of 
matter \cite{fisher79}. Different sorts of critical points represent different sorts of states, and each has its own
  particularness and peculiarities. 

It is in this vein that the current article discusses {\it fluctuation-dominated phase ordering}  (FDPO).  Like the critical points discussed above, FDPO often arises at the separatrix between ordered and disordered phases in a number of equilibrium and non-equilibrium systems. But it differs from a normal critical state in a fundamental way. As explained in the sections that follow, the state is characterized by long-range order whose magnitude varies strongly owing to fluctuations; hence the name “fluctuation-dominated order”. 

In general, there are two properties which characterize FDPO. The first is that the distribution of the order parameter remains broad, even in the thermodynamic limit. The second is that the two-point spin-spin or density-density correlation function depends on separation scaled by system size, and the scaling function exhibits a cusp singularity. These points are discussed in the subsections below. First identified in a nonequilibrium system of passively sliding particles on a fluctuating surface \cite{dib2000,dib2001}, FDPO has been found in several other nonequilibrium systems, as well as an equilibrium Ising model with long-range interactions \cite{mbtidsi2019}.

\subsection{Broad distribution of the order parameter}

In a typical phase ordering system which exhibits two phases, a large finite system in steady state resides in one phase for a  long time before switching over to the other. For instance, in an Ising ferromagnet if  $s_j$ is the spin at site $j$, then the order parameter with non-conserved dynamics is  $m = {\frac{1}{L}} \sum_{j=1}^{L} s_j$. The probability distribution of $m$ is sharply peaked around the two values $m_s$ or $-m_s$, with very infrequent transitions between the two. The peak widths around $\pm m_s$ approach zero in the thermodynamic limit $L \rightarrow \infty$.

By contrast, in the FDPO steady state, the system shows large fluctuations without losing macroscopic order. Thus the order parameter distribution remains broad:
\begin{equation}
Lim_{L \to \infty}P(m) =  f(m)~~~~~~~~~~~~~~~~~~~~~~~~~~~~[FDPO]
\end{equation}
The limiting distribution $f(m)$ does not narrow down, and  has support in a broad range of $m$, even in the thermodynamic limit.

 If the order parameter is conserved, the magnetization is constant in both ordered and disordered phases. Quantities which are sensitive to the difference between phases are the long-wavelength Fourier modes $Q_k = {\frac{1}{L}} \sum_{j=1}^{L} |e^{iqj} (1+s_j)/2| $, with  $q = \frac{2 \pi k}{L}$ and $k$ a small integer.   For a given conserved value of magnetization $m$, the mode $Q_1$ is an appropriate order parameter. In an Ising ferromagnet, the probability distribution of $Q_1$ is sharply peaked at a single value. But in the FDPO state with a conserved order parameter, the distribution of $Q_1$ remains broad in the thermodynamic limit, as do the distributions of $Q_2, Q_3$ ... \cite{kapri2016}.

Another important characteristic of the order is obtained from the asymptotic value of the two point spin-spin correlation function $C(r) = < s_0 s_r >$ in the absence of a field. At large separations $r$, the correlation function is expected to decouple
  \begin{equation}
Lim_{r \rightarrow \infty} Lim_{L \rightarrow \infty} <s_0 s_{0+r}> = <s_0> < s_{0+r}> = {m_c}^2
\end{equation}
A nonzero value of $m_c$ indicates that the system has long-range order. In a finite system, $C$ is a function of the scaled variable $r/L$ for large enough values of $r$ and $L$. The value ${m_c}^2$ can be read off as the intercept in a plot of $C$ versus $r/L$. While $m_c$ and $m_s$ coincide in systems such as the ferromagnetic Ising model, this is typically not the case with FDPO. 

\subsection{Singularity in the scaled correlation function}

As discussed above, FDPO often arises at the separatrix between ordered and disordered phases. Normal critical behaviour is characterized by a power law decay  $C(r) \sim r^{d - 2 + \eta}$ where $\eta$ is the Fisher exponent; by contrast,  in the FDPO state, the correlation function is a function of $r/L$ rather than $r$ alone. It shares some features with phase ordering systems (scaling as a function of $r/L$), and other features with critical systems (strong fluctuations, power law decays, singular scaling functions).

To appreciate the significance of the cusp singularity in $C$, first let us recall some facts about phase ordering in a system with regular ferromagnetic Ising order. When  the order parameter is conserved, the low-$T$ state is phase separated, with each phase occupying a macroscopic region. An interface of width $W$ separates the phases, where $W$ is of the order of correlation length $\xi$. Thus on length scales $\xi << r <<L$, the interface is structureless and sharp. If further we consider $r/L << 1$, then $C(r)$ takes the form $C(r) \approx {m_c}^2 (1 - 2r/L)$. The linear fall can be understood as follows: spatial averaging of  $s_o s_r$ produces  $+{m_c}^2$ with probability $(1- r/L)$ if points $0$ and $r$ lie within a phase, and $-{m_c}^2$ if they lie across an interface. Further, the linear drop in $r/L$ implies that the structure factor $S(q)$ (the Fourier transform of  $C(r)$) is given for large scaled wave-vectors $qL >> 1$ by $S(q)/L^d \sim 1/{(qL)}^{d+1}$. This form of the decay of $S(q)$ is known as the Porod Law. The Porod Law is well established in scalar order parameter systems with a sharp interface, and is widely observed \cite{porod, bray}.

\begin{figure}[ht]
\centerline{\includegraphics[width=7cm]{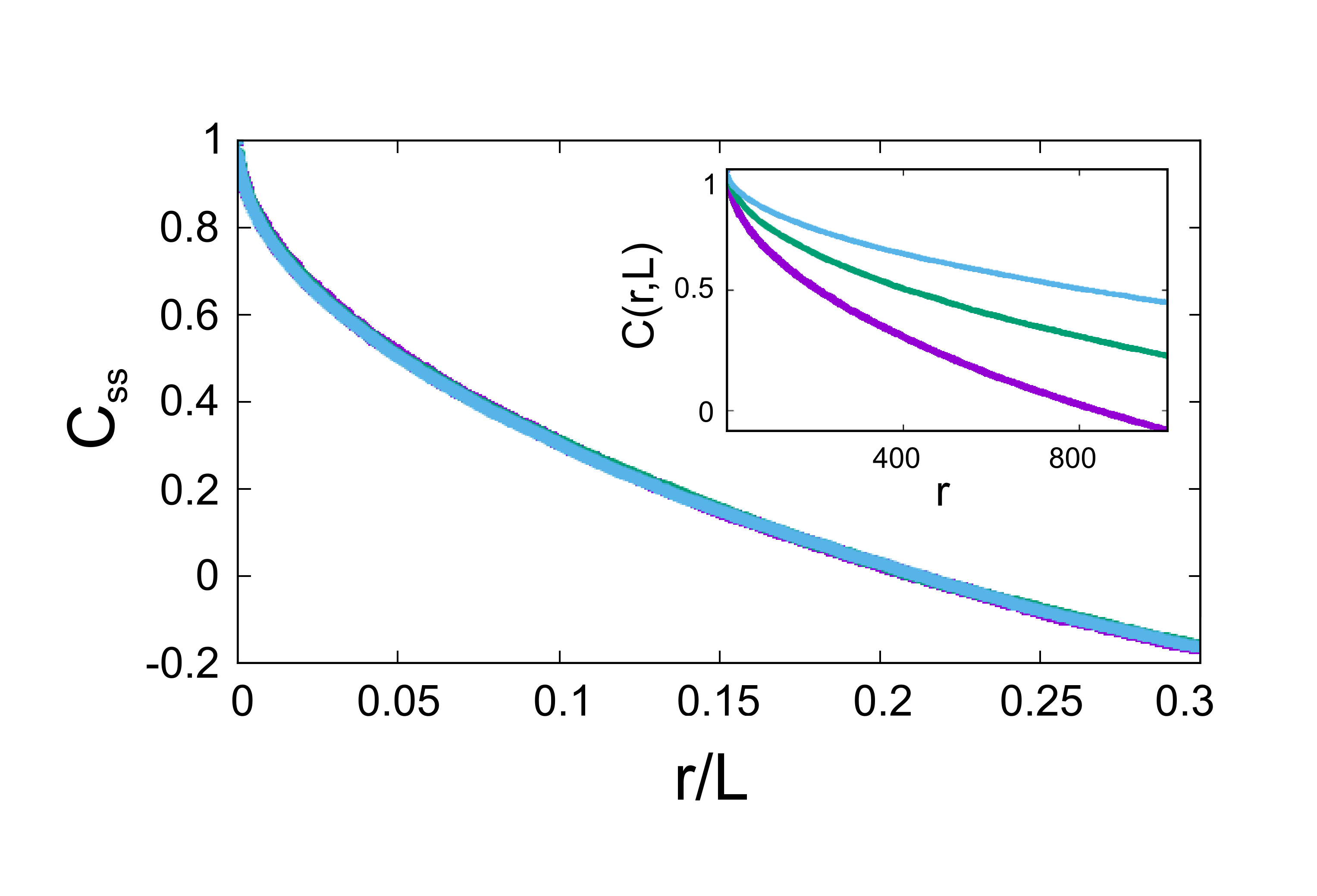}
\includegraphics[width=7cm]{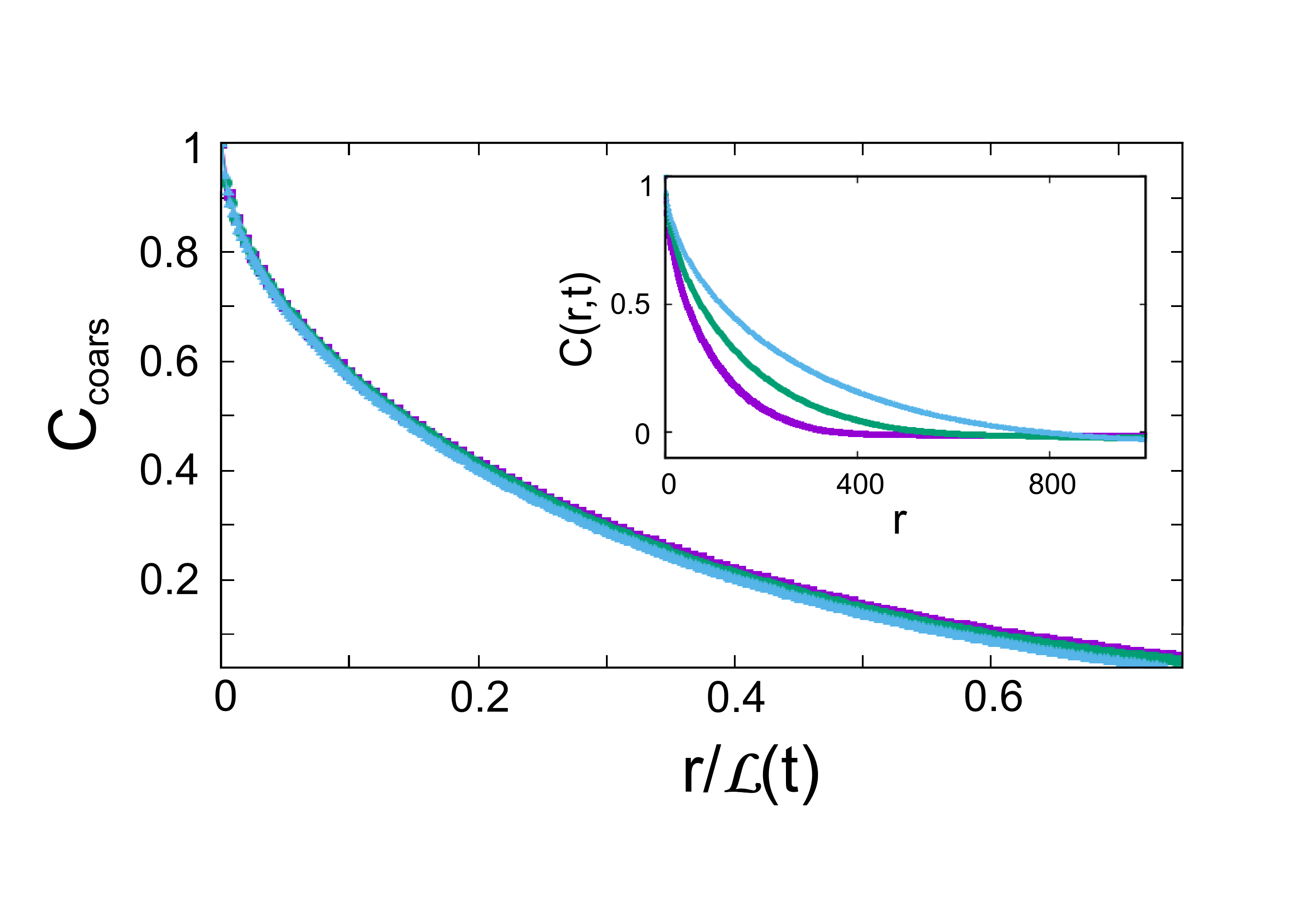}
}
\caption{In most systems which exhibit FDPO, the two-point correlation function is a singular function of the scaled separation. The data shown is for the coarse-grained depth (CD) model, which is defined in Section 3. In the left panel, the separation $r$ is scaled by the system size $L$ in steady state (with $L=4000, 8000, 16000$), while the right panel shows scaling by a time-dependent length scale ${\cal L}(t)$ while coarsening (with ${\cal L}(t) = 68, 100, 147$). In both cases, the scaling function exhibits a cusp singularity at the origin, indicating that the Porod law beaks down. The unscaled data is shown in the insets.}  
\label{fig7}
\end{figure}
 
A prominent signature, found in most systems that exhibit  FDPO, is the singular behaviour of the two-point correlation function $C(r,L)$ between points separated by a distance $r$ in the steady state of a system of size $L$. As for a regular ordered state, the data shows a scaling collapse when plotted against the ratio $r/L$, and the steady state  scaling function $C_{ss}$ has a finite intercept at the origin (Fig. 2). The intercept gives the value of long range order in the system, as the ratio $r/L$ is driven to zero as $L \rightarrow \infty$ for every  value of the separation $r$. The key distinguishing feature of FDPO is the ‘cuspy’  approach to $r/L = 0$, seen in Fig. 2. The steady state scaling function shows a cusp singularity:
\begin{equation}
C_{ss} \approx {m_0}^2 - b_{ss} \left({\frac{r}{L}}\right)^{\alpha} ~~~~~ (r/L \rightarrow 0) ~~~~~~~~~~~~~~~~~~~[FDPO]
\end{equation}
where the cusp exponent $\alpha < 1$. This feature implies the breakdown of the Porod Law discussed above,  according to which the intercept is approached linearly ($ \alpha =1$). As argued above, the Porod Law relies only on the existence of a well-defined interface  separating coexisting phases, with the interface width being finite. That the law no longer holds implies that in FDPO, the region between coexisting ordered phases is also proportional to $L$, and occupies a  fluctuating but finite fraction of the system. This is illustrated in Fig. 3 which shows a typical configuration of the $CD$ model in steady state. Two features stand out: first, the occurrence of some very long (horizontal) stretches of like spins; and second, that these long stretches are separated from each other by a large disordered region where spins are scrambled up, owing to a conglomeration of close-by domain walls. The longest stretches of like spins are of order system size $L$, consistent with the occurrence of long range order. On length scales set by $L$, the scrambled-spin regions are disordered, and the largest of these are associated with the departure from the Porod law.

\begin{figure}[ht]
\centerline{\includegraphics[width=11cm]{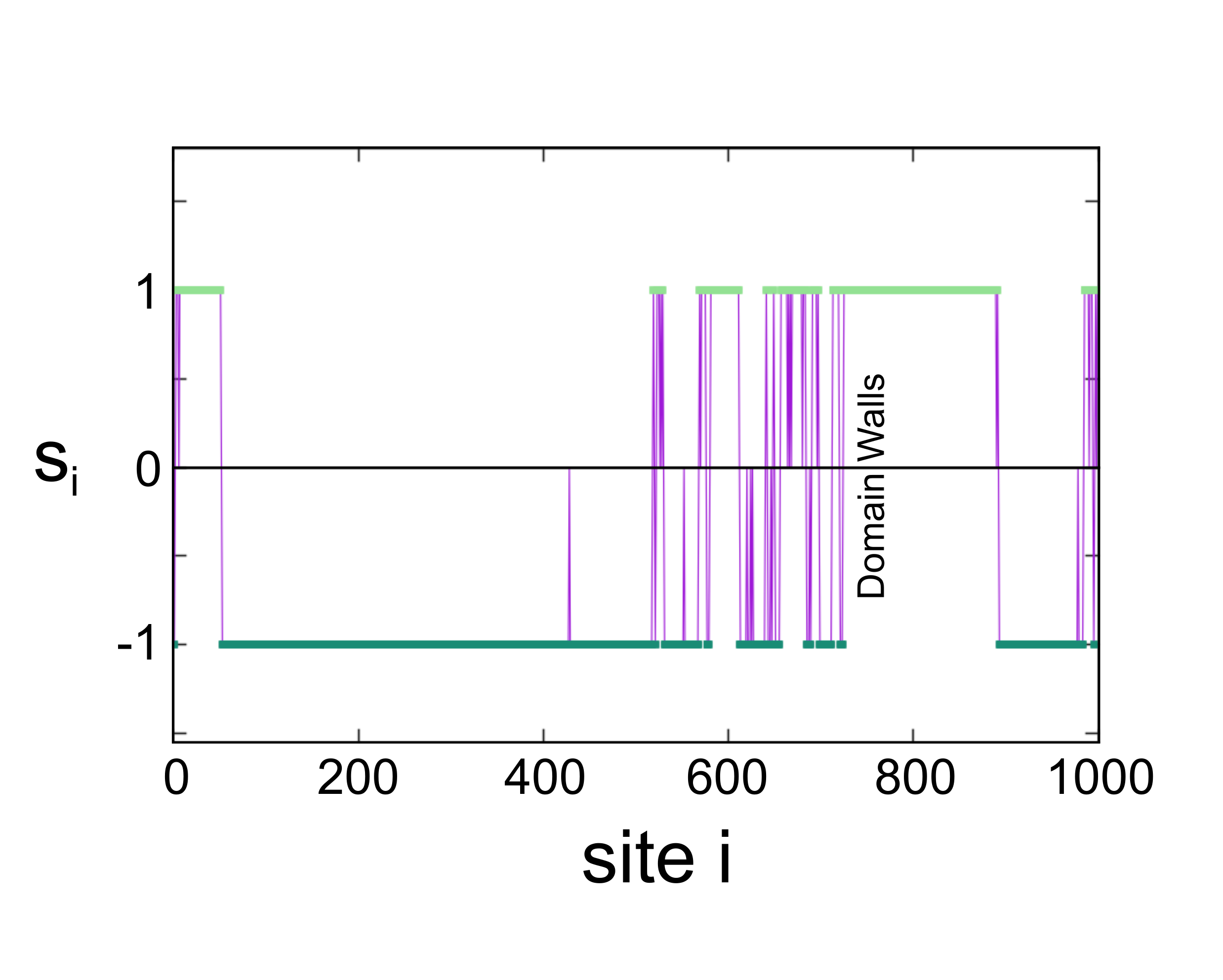}}
\caption{A typical configuration in the steady state of the coarse-grained depth model defined in Section 3. The horizontal lines depict regions of like spins, while the vertical lines are domain walls. Macroscopic stretches of like spins are responsible for long-range order, while spins are scrambled in disordered regions with many domain walls. Large disordered regions are responsible for the breakdown of the Porod law.}
\label{fig8}
\end{figure}

A similar situation arises also while the system is coarsening. Consider an infinite system evolving in time, starting from a completely disordered state. As known from studies of phase ordering dynamics \cite{bray},  the system reaches a local quasi-steady state on length scales smaller than $ {\cal L} (t) \sim t^{1/z}$ where $z$ is the dynamic exponent. In particular, we expect the analogue of Eq. (1) to hold, with system size $L$ replaced by the length scale $ {\cal L} (t) $. The right panel of Fig. 2 shows that this is indeed the case, with the coarsening data showing that the scaling function $C_{coars}$ has a cusp as a function of $r/{\cal L}(t)$. 


\subsection{Systems which exhibit FDPO}

Most systems which show FDPO show both a broad order parameter distribution and a singular scaled correlation function. In exceptional cases, one of these may be suppressed \cite{chandrashekar2023}, but we will not consider such cases below. We summarize the content and principal results of the sections that follow.

Three simple models which exhibit FDPO are discussed in Section 3. The first is a model of particles sliding on a fluctuating surface, where  numerical studies establish FDPO in the system. Next, motivated by the extreme adiabatic limit of this model, we study the coarse grained depth (CD) model of the fluctuating surface.  Both the broad order parameter distribution and the singularity in the scaling function can be established analytically in this case. Finally, we study the truncated inverse distance squared Ising (TIDSI) model, an Ising model with long range interactions that act within a cluster of like spins. Along a part of the critical locus in this model, the system shows FDPO, and the cusp exponent can be determined analytically. Interestingly, it is found to  depend continuously on the coupling constants. This model is also discussed in the article on mixed order phase transitions in this volume \cite{mukamel2023}.

In Section 4, several issues related to FDPO are discussed, including the imprint of FDPO on time-dependent properties. The autocorrelation function in the steady state of the model shows the analogue (in the time domain) of the cuspy behaviour discussed above for the equal-time two point correlation function. Further, while the system is coarsening, it shows an interesting aging behaviour which is captured by a two-time correlation function.

Also discussed in Section 4 is {\it fluctuation-dominated clustering} (FDC), which generalizes FDPO to describe situations with much more intense clustering. FDC is accompanied by  even larger fluctuations, and the cusp singularity of Eq. (3) is replaced by a divergence in this case. Finally, several systems which display FDPO and FDC are discussed, including passive particles carried by a compressible fluid, lipid clustering in cell membranes, coarsening in the Vicsek model, active nematics, vibrated rods and inelastically colliding particles.


\section{Models of FDPO}

\subsection{Particles sliding on a fluctuating surface}

Consider a system of particles which lie on a fluctuating surface and slide downward on it (Fig. 3). The time evolution of the surface is assumed to be governed by either Edwards Wilkinson (EW) or Kardar Parisi Zhang (KPZ) dynamics \cite{ew,kpz}. The particles are guided downward by the surface gradient, but are passive in that they do not influence the time evolution of the surface. While the sliding particle (SP) model can be  defined and studied in any dimension, many of the most interesting features are seen in one dimension. Results for the  two-dimensional case are discussed at the end of this section. 

In a lattice representation of the one-dimensional model, particles (which are taken to have a hard core interaction which enforces exclusion) occupy the sites of a 1D lattice with periodic boundary conditions whose bonds represent the local slopes of the surface. The surface dynamics is that of the single-step model in which corners are flipped stochastically \cite{plischke}, i.e. $ \wedge \rightarrow \vee $ with rate $p_1$, while the reverse transition $ \vee \rightarrow \wedge $  occurs with rate $q_1 \le p_1$ (Fig. 4). Equal rates $q_1 = p_1$ correspond to EW dynamics, while unequal rates lead to KPZ surface dynamics. The rate of particle-hole exchanges on nearest neighbour sites depends on the state of the bond in between. If the exchange across a bond results in a lower location of the particle, then the move occurs with rate $p_2$, while the reverse transition occurs with rate $q_2 < p _2$. 


\begin{figure}[ht]
\centerline{\includegraphics[width=11cm]{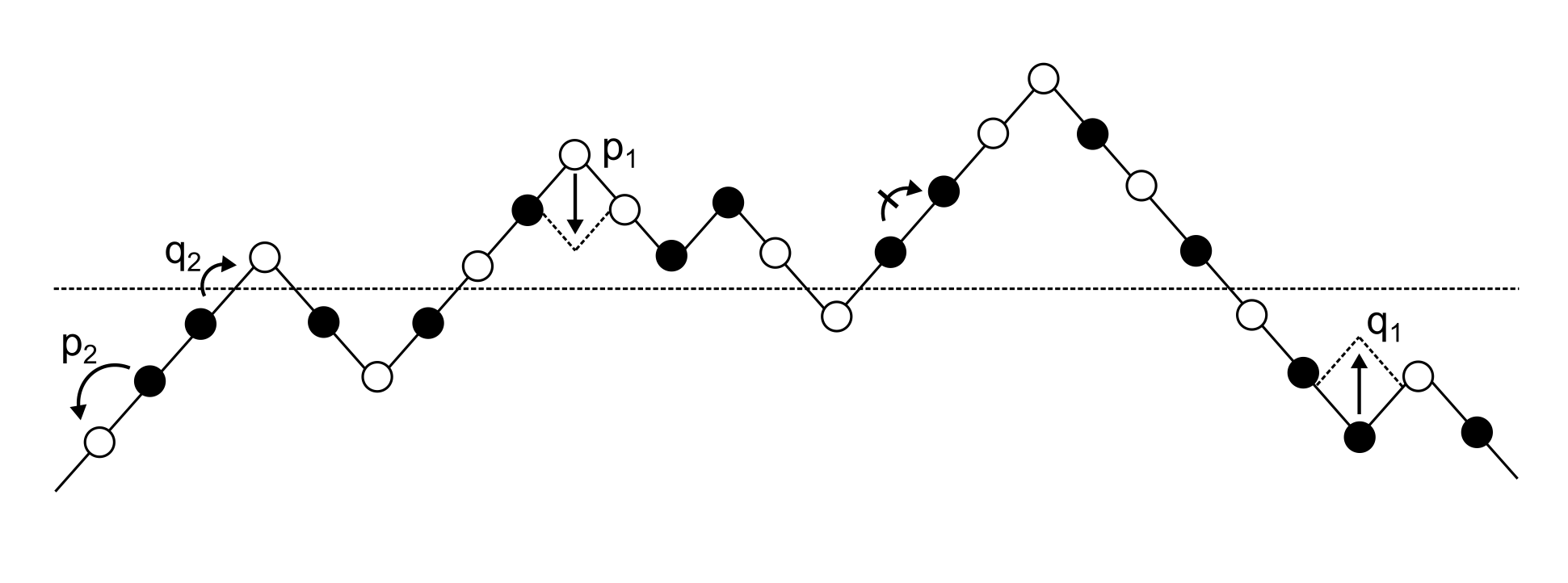}}
\caption{The sliding particle (SP) model describes passive particles sliding on a fluctuating surface. The surface moves involve a local hill becoming a local valley, and vice versa, while a particle can hop onto a neighboring vacant site, with downward hops occurring at a greater rate. Also shown is a horizontal cut, required for the definition of the coarse-grained depth (CD) model, in which the state on site are decided by the height of that site with respect to the cut.}
\label{fig8}
\end{figure}

Surface fluctuations lead to particle clustering. The reason is that particles easily slide down a surface fluctuation which creates a valley; but, should the reverse surface fluctuation occur, it is very difficult for particles to climb up a hill as the process involves the slow process of activation against a field. Given this propensity to promote particle clustering, it is interesting to ask about the large-scale behaviour as time increases. 

Starting from an initially random placement of particles on the surface, the evolution of the local density
 can be tracked by numerical simulation. In a system with $N_P$ particles, particularly revealing is the evolution of the two-point  density-density correlation function  $C(r,t) \equiv <n_i n_{i+r}> -(N_P/L)^2$. It is found that the data at different times collapse, on scaling the separation by a growing length scale ${\cal L} (t) \sim t^{1/z}$, much as in the right panel of Fig. 2 which refers to the coarse-grained depth (CD) model defined in the next subsection. This is expected, as catchment areas for particles are provided by fresh valleys which form when coarse-grained hills overturn, and the bases of such hills grow as  $t^{1/z}$ where $z$ is the dynamic exponent which governs surface fluctuations ($z=2$ for EW evolution, $z=3/2$ for KPZ evolution). In the scaling limit, the data collapses to a single function $C_{coars} (y=r/{\cal L}(t))$. The intercept indicates the value of long-range order ${m_c}^2$. The approach to the intercept exhibits a cusp singularity $C_{coars}(y) = {m_c}^2 - b y^{\alpha}$, which signals the breakdown of the Porod law, according to which  $C_{coars}$ should fall linearly.

In a finite system of length $L$, the system reaches the steady state in a time of order $L^z$. In steady
state, the two-point correlation function $C(r,L)$ varies with separation $r$ on the scale of system
size $L$, and the data for different sizes $L$ shows a collapse when plotted as a function of $r/L$, in
 the scaling limit $r \rightarrow  \infty, L \rightarrow\infty$ with $r/L$ held fixed (as in the left panel of Fig. 2, which refers to the CD model). The scaling function $C_{ss}$ for steady state correlations shows a cusp singularity as $y=r/L$ approaches zero, $C_{ss}(y) = {m_0}^2 - b y^{\alpha}$, pointing to FDPO. This is to be contrasted with customary phase ordering, where $C(r) = m_c^2 (1- 2 r/L)$ for $r/L<<1$, on length scales $r$ much larger than the correlation length $\xi$, which determines the width of the interface.

An interesting point is that the SP model actually lies on the phase boundary separating a fully ordered state from a disordered state. With EW driving, this is the phase boundary of the Lahiri-Ramaswamy model, a lattice model of sedimenting colloidal crystals \cite{lahiri2000}, while with KPZ driving, this is the phase boundary separating disordered and ordered states of the light-heavy (LH) model \cite{shauri2017,samvit2020}. In both cases, the SP model shows FDPO, while the order parameter shows a first order 0-1 jump from the disordered to the ordered phase. This is consistent with the transition being of mixed order, but in order to clinch the mixed order character (see Section 3.3), the divergence of the correlation length on approaching the critical locus from the disordered side needs to be shown for SP model.
 
In two dimensions, a study of hard core particles sliding down a surface evolving through KPZ dynamics 
shows that particles cluster together on a length scale which grows in time \cite{manoj2003}. The final fluctuation-dominated  phase-separated steady state is characterized by a power-law distribution of cluster sizes of particles and holes, though the two are not identical, unlike in 1D.  The scaled correlation function shows a cusp singularity and the order parameter distribution remains broad in the large-size limit. We conclude that FDPO is present in this case as well.
 
 \subsection{The Coarse-grained Depth Model}
 
 Michael Fisher always emphasized the importance of simple models, preferably those that can be analyzed analytically, to shed light on the phenomenon of interest. In the context of FDPO, this role is played by the Coarse-grained Depth (CD) model of a fluctuating surface \cite{dib2000, dib2001, ad2023}.
 
 The CD model is motivated by considering the extreme adiabatic limit of the sliding particle (SP) model considered above. In this limit, particles tend towards the minimum energy configuration for a particular height profile. Since the particles have hard core interactions, this state corresponds to filling the height profile, bottom upwards; the topmost particles define a horizontal level below which all sites are occupied by particles. 
 
 We define the CD model by taking a horizontal cut on a fluctuating surface with zero net slope (Fig. 4). For a given configuration of the surface, if the cut is at height $H(t)$,  we define a spin  variable at site $i$ as $s_i(t) = - sgn [h_i(t) - H(t)]$. Evidently $s_i(t)$  can take values +1, -1 or 0  depending on whether the height at the $i$'th site is below, above or at the cut. It provides a coarse-grained description of the surface height. The time evolution of  $s_i$ is induced by the dynamics of the underlying surface. Invoking the correspondence with the adiabatic SP model, sites with $s_i = 1$ would correspond to occupied sites.  However the correspondence is not exact, as $\sum s_i $ is not conserved by the dynamics. 
 
Different choices of the reference height $H(t)$ correspond to different variants of the CD model \cite{dib2001, ad2023}. Two useful choices are (i) The {\it CD2 model}, in which the reference level is chosen to be at the instantaneous height of the site $i=0$, i.e. $H(t) = h_{i=0}(t)$. This choice leads to a useful mapping to random walks. (ii) The {\it CD3 model}, in which the reference level is chosen to be at the instantaneous mean height $H(t) = <h(t)>= \frac{1}{L} \sum h_i (t) $. 

There is a simple mapping between each surface configuration in the steady state of the CD2 model and a random walk (RW) trajectory, with a  site label in the former representing time in  the latter,  and the local surface slope  $ +1$ or -1 representing a rightward or leftward RW step. Several results follow immediately from the mapping. (a) The length of a cluster of like spins maps to the time between successive returns to the origin, implying that the distribution of cluster lengths $P(l)$ in steady state $\sim l^{-3/2}$ for $1<<l<L$.   (b) For the CD2 model with periodic boundary conditions, the problem is isomorphic to the Brownian bridge \cite{godreche}. (c) As the longest return time for a $T$-step RW is of order $T$,  it follows that  that  the length $l_{max}$ of the longest cluster in the CD2 model is of order $L$. The distribution of the longest cluster $P(l_{max})$ is a function of the scaled variable $l_{max}/L$ \cite{godreche, frachebourg}.Thus both the mean and standard deviation of $l_{max}$ are of  $O(L)$.  (d) The order parameter  $\tilde{m}=\frac{1}{L} \sum s_i$ maps to the excess time a random walker spends on one side of the origin. This distribution is known to be uniform, \cite{feller} implying that the order parameter distribution in Eq. (1) is independent of $\tilde{m}$.  (e) Knowing $P(l)$ allows the two-point correlation function to be calculated within the independent interval approximation \cite{dib2000,dib2001}. The result is that $C(r,L)$ is a scaling function of $y=r/L$, which, for small argument, reduces to Eq. (2) with the cusp exponent $\alpha = 1/2$ \cite{dib2001}. 
We conclude that the CD2 model exhibits the hallmarks of FDPO: a broad distribution of the order parameter, and a cusp singularity in the scaled correlation function. These features are confirmed  by numerical simulation for the CD3 model as well \cite{ad2023}. For instance, Fig. 2 shows that $C(r,L)$ is a scaling function of $r/L$ with a cusp singularity for small argument. 

We expect these steady state results for the CD models to generalize to the coarsening regime, with the coarsening length scale ${\cal L}(t)$ taking the place of  $L$. Indeed, the  cluster size distribution shows a power law decay $P(l) \sim l^{-3/2}$  with an exponential cut-off at $l \sim {\cal L}(t)$ \cite{ad2023}. Further,  as shown in the right panel of Fig. 2, the two-point correlation function of the CD3 model during coarsening is a scaling function of  $r/{\cal L}(t)$, with a cusp singularity at small argument. 

Finally we remark that the CD model can be generalized to higher dimensions by taking a cut of an
 evolving self-affine surface. On a 2D KPZ surface, there is a cusp singularity in the scaled correlation function and a broad distribution of the order parameter, pointing to FDPO \cite{manoj2003}.
 
 \subsection{FDPO at a mixed order transition: Ising model with long-range interactions}
 
 Fluctuation-dominated order also occurs in an equilibrium system, namely a 1D Ising model with truncated power-law decaying interactions \cite{mbtidsi2019}.  This model shows  a {\it mixed-order transition}, characterized by a discontinuity of the order parameter across a critical locus, along with a divergent correlation length from the disordered side; FDPO occurs along part of the critical locus (Fig. 4). 

In the truncated inverse-square decay Ising (TIDSI) model, a nearest neighbour interaction  $-J_{NN}\sigma_j\sigma_{j+1}$ is present, along with a long-range interaction term $ -J(i-j) \sigma_i\sigma_j I(i~j)$ where $ I(i~j) = 1$ if sites $i$ and $j$ belong to the same domain of all up or all down spins \cite{bar2014}. The coupling $J(r)$ is taken to be $J( r) \approx J_{LR}/r^2$. The Hamiltonian is thus

\begin{equation}
\mathcal{H} =-J_{NN} \sum_{i=1}^{L} \sigma_i \sigma_{i+1} - 
\sum_{i<j} J(i-j)   \sigma_i \sigma_j \prod^{j}_{k=i} 
\frac{1+\sigma_k \sigma_{k+1}}{2}
\end{equation}

The analysis of the model is described in the article on mixed order transitions which appears in this volume \cite{mukamel2023}, and thus we give only a brief account below. It is convenient to go to a domain representation, in which a configuration is described by a set of domains $(l_1, l_2, ...l_N)$ with each domain being a stretch of parallel spins. The number of domains $N$ can vary from one configuration to another. In the domain representation, the Hamiltonian is rewritten in terms of $\{ l_i \}$, and the partition function can be expressed as
\begin{equation}        
Z_y(L) =  \sum^\infty_{N=1} y_0^N \sum^\infty_{l_1=1} 
\ldots \sum_{l_N=1}^{\infty} \prod^N_{n=1} 
\frac{1}{l_n^c} \, \, \, \delta_{\sum^N_{n=1} l_n, \,L} \, .
\label{pf.1}
\end{equation}
where  $y_0=\exp{-\beta \Delta}$ is the fugacity of a domain wall and $c=\beta J_{LR}$.  Here $\beta$ is the inverse temperature while $\Delta = 2 J_{NN} + J_{LR} + b_0$. 
The analysis of the partition function as a function of the coupling constants $c$ and $y_0$  leads to the phase diagram shown in Fig. 5, with the critical locus given by the equation $y_0 = 1/\zeta(c)$ where $\zeta$ is the Riemann zeta function. Below we discuss the behaviour in the disordered and  ordered phases, and along the critical locus \cite{mbtidsi2019}.
\begin{figure}[ht]
\centerline{\includegraphics[width=11cm]{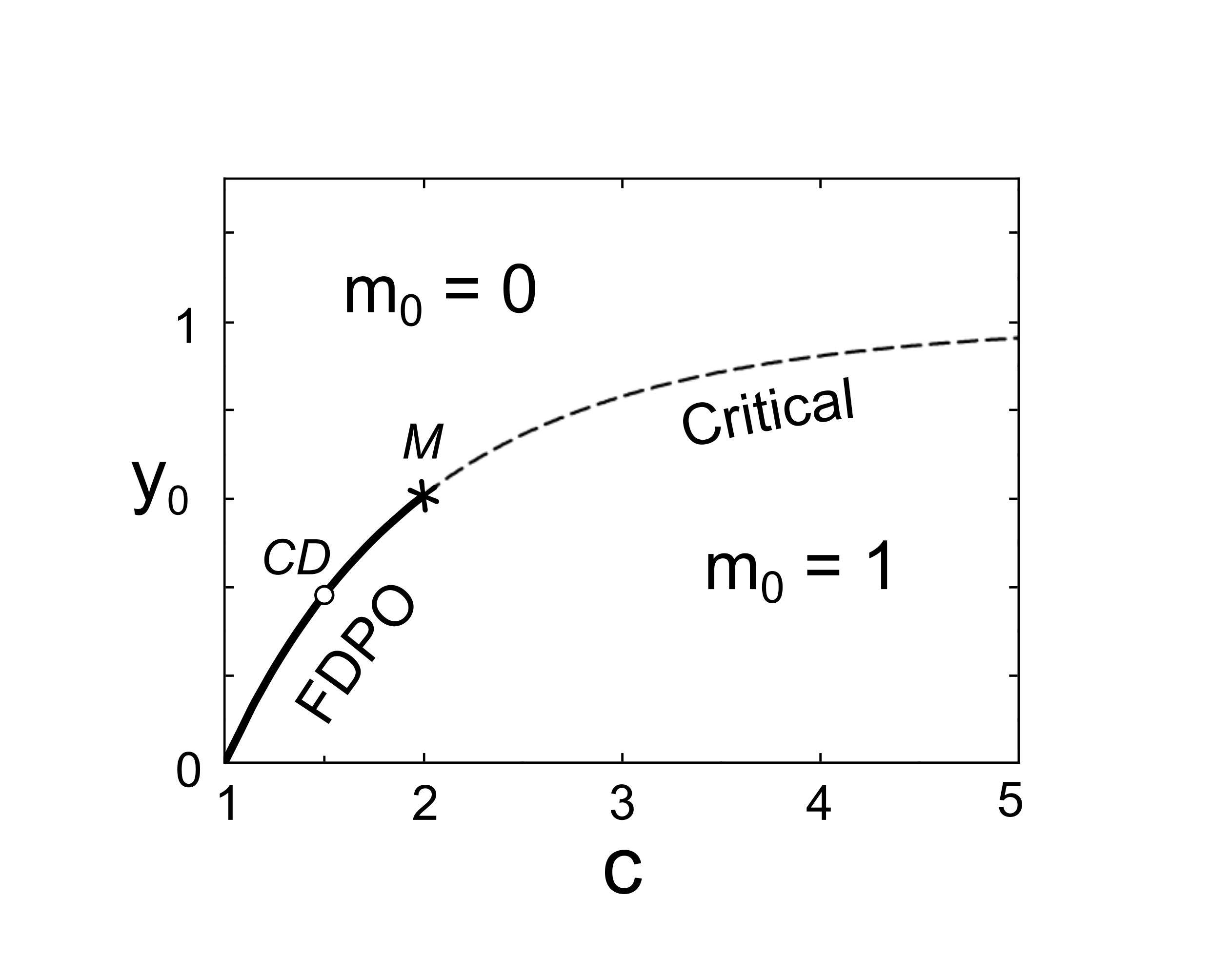}}
\caption{Phase diagram of the TIDSI model in the $(c-y_0)$ plane, with $c=\beta J_{LR}>1$. The critical line separates the paramagnetic $m_0=0$ phase from the ferromagnetic $m_0=1$ phase. The portion of the critical line with $1<c<2$ exhibits FDPO. The multicritical point $M$ marks the change from FDPO to normal critical behaviour, while at the point marked CD, the large-distance behaviour is the same as that of the CD model. (Adapted from {J. Phys. A: Math. Theor}, 2019. \cite{mbtidsi2019})
}
\label{fig8}
\end{figure}

In the disordered phase, the net magnetization $m_0 = \frac{1}{L} \sum <s_i>$ vanishes. The probability distribution of domain lengths has an exponential cutoff, $p(l) \sim exp(-l/\xi)$ which serves to define the correlation length $\xi$. As the critical line is approached, the domain wall density approaches zero  and  $\xi \sim t ^ {-\nu}$ where $t = (T -T_c) / T_c$ is the reduced temperature; the value of the critical exponent is $\nu =1$ for $c > 2$  and $\nu = 1/(c-1)$ for $ 1< c < 2$.

In the ordered phase, the value of $m_0$ is 1, as the fraction of sites in the longest cluster approaches unity in the limit $L \rightarrow \infty$ \cite{baretal2016}; the state is one with near-perfect order. Thus
$m_0$ shows a 0-1 jump, while $\xi \rightarrow \infty$ in the disordered phase,  implying a mixed order phase transition across the critical line.

Along the critical locus, the system exhibits two distinct behaviours \cite{mbtidsi2019}. For $c > 2$, the system shows normal critical behaviour, whereas it exhibits FDPO in the range $1 < c < 2$.  In the former case, the mean cluster size $<l>$ is finite, and the mean size of the largest cluster is a vanishing fraction of the system size as $L \rightarrow \infty$. The correlation function $C(r,L)$ decays asymptotically as a simple power law $\sim r^{- (c-2)}$.  On the other hand, for  $1 < c <2$,  the mean size of the largest cluster $\sim L$ \cite{baretal2016}. The correlation function can be found from the cluster size distribution within the independent interval approximation.
The result is $C(r,L) \approx 1 - (c-1)^{-1} \left(\frac{r}{L}\right)^{2-c}\, $ 
which is of the FDPO form of Eq. (1). Interestingly, the cusp exponent $\alpha$ depends continuously on the parameter $c = \beta J_{LR}$.

These results for the steady state suggest that a similar form would hold for the TIDSI model while coarsening, with system size $L$ replaced by the coarsening length ${\cal L}(t)$. This has been checked by numerical simulation using single spin-flip (Glauber) moves for the dynamics \cite{souviksaroj}.

The domain representation of the probability of a configuration brings out an interesting connection between the TIDSI and CD models, namely that the steady state properties of the CD model correspond to those of the critical TIDSI model with c=3/2 \cite{ad2023}. 
However, this correspondence for static quantities does not extend to the dynamics, as the interface-driven dynamics of the CD model is quite different from the single-spin flip dynamics in the TIDSI model.



 \section{Discussion and Conclusion}

Below we discuss several issues concerning FDPO and related states. In turn, we discuss the distinctive imprint of FDPO on dynamic properties;  a generalization of FDPO in which clustering is much more intense, namely fluctuation-dominated clustering (FDC); the occurrence of non-Porod behaviour in systems with quenched disorder; and finally, examples of several physical phenomena and models in which fluctuation-dominated states are key.

\subsection{Dynamic manifestations of FDPO}
 
Besides its effect  on spatial correlation functions, FDPO has a strong effect on dynamics. This is true both for the autocorrelation function in steady state, and for aging correlation functions during coarsening for both SP and CD models \cite{sakuntala2006}. Interestingly, the cuspy behaviour of scaled spatial correlation functions is inherited by dynamic correlations.

We define the autocorrelation function  as  $A(t) = <s_i(0)s_i(t)>$. It is found that $A(t,L)$ is a scaling function of $t/L^z$ where $z$ is the dynamic exponent governing the surface dynamics. For $t/L^z<<1$, we find
\begin{equation}
A(t,L) \approx {m_c}^2 - a \left({\frac{t}{L^z}}\right)^{\beta '}  .
\end{equation}
In the cases considered, we find that the exponent  $\beta’$  is smaller than 1, implying that cusp singularities in the scaling function are to be found in dynamics as well.  In the case of the CD model,
$A(t,L)$ can be calculated analytically for the CD3 model driven by an EW surface\cite{sakuntala2006}. One finds that  Eq. (6) holds, with ${m_c}^2 =1, \beta’ = 1/4, z=2$.
With KPZ evolution for the surface, Monte Carlo simulation results  for $A_{CD}(t)$ show that Eq. (6) holds with $z=3/2$ and the cusp exponent  $\beta' \simeq 0.31$. For the SP models too, numerical simulations confirm the validity of Eq. (6). Finally, the two-time correlation function which characterizes aging dynamics in the coarsening regime can be calculated analytically in all regimes for the CD3 model with EW driving, and determined numerically in other cases \cite{sakuntala2006}.

\subsection {More intense clustering: Fluctuation-dominated Clustering (FDC)}

In FDPO, the two-point correlation function exhibits scaling when separations are scaled by $L$ in steady state (or by ${\cal L}(t)$ during coarsening). In either case, the scaling function exhibits a cusp singularity in the limit that the scaled separation $y=r/L$ (or $r/{\cal L}(t)$) approaches zero. 

However, a cusp is not the only possibility \cite{barma2008}. In a number of systems, $C(r,L)$ shows a stronger singularity, namely a divergence:
\begin{equation}
C(r,L) \approx L^{-\mu}Y \left({\frac{r}{L}}\right) 
\end{equation}
with
\begin{equation}
~~~~~~~~~~~~~~~~~Y(y) \sim  y^{-\nu} ~~~~~~~~~  (y \rightarrow 0)~~~~~~~~~~~[FDC]
\end{equation}
The divergence indicates much stronger clustering. It embodies two effects: first, the formation of  clusters, each made of a large number of particles; and second, the clustering of clusters close to each other. The clusters are dynamic entities which keep changing in time, as they make and break. Correspondingly, the state is accompanied by large fluctuations and exhibits spatial and temporal intermittency, {\it i.e.} it shows {\it fluctuation-dominated clustering} (FDC).  

In the context of particles sliding on a fluctuating surface, FDC arises when the particles are noninteracting. Numerical simulations show that particles move towards valley bottoms, and since there is no restriction on the number $n_i$ of particles on each site $i$, the degree of  clustering is much larger than in FDPO \cite{apoorva2005}. Parenthetically we note that the ``strong clustering states" discussed in this reference are identical to the FDC states under discussion here. The divergence in Eq. 8 is strongest in the case of a KPZ surface, in which case $\nu=3/2$ and $\mu=1/2$. With driving by an EW surface, Eqs. 8 and 9 hold with $\mu=0$ and $\nu \simeq 2/3$. In both cases, there are strong signatures of intermittency, in space and time \cite{tapas2018}.

The scaling function can be found analytically in the adiabatic limit when the surface profile is stationary; the problem reduces to the Sinai problem of random walkers in a random medium, which itself is the trajectory of a random walk. In an analytic treatment of the steady state of the model, the site-wise distribution of particle numbers was derived, as well as the two-point correlation function which was shown to be of the FDC scaling form (Eqs. 7 and 8) \cite{apoorva2005, apoorva2006}. The     results agree well with numerical results for noninteracting particles sliding down a KPZ surface \cite{apoorva2006}, and are consistent with earlier work on random walkers in a Sinai landscape \cite{comtet}.

We note an interesting point regarding the connection between giant number fluctuations and the form of the scaling function \cite{supravat2012}. While singularities in the small $y \equiv r/L$ behaviour (cusps for FDPO, divergences for FDC) indicate a departure from the Porod Law, the correlation function may, in some cases, exhibit a  different power law decay for large values of $y$. Now fluctuations of the total number of particles in a given volume involve a sum over correlation functions in that region, and this sum is dominated by the contribution from large separations. In such cases it is the large-$y$ power law in the scaling function which is pertinent to determining number fluctuations \cite{supravat2012}.

\subsection{Disordered systems which show non-Porod behaviour} 

Correlation functions in some systems with quenched disorder show strong deviations from the Porod law. It is useful to distinguish between two classes of quenched systems. In the first class, we are interested directly in the correlation functions of the quenched degrees of freedom, while in the second class, the quenched disorder provides a background for other degrees of freedom, and we are interested in the correlations in the latter set.

In the first class, the degrees of freedom are frozen. An important example of this class is porous materials, in which two-point correlation functions have been studied extensively by scattering techniques, and non-Porod behaviour has been established \cite{bale1984, wong1985}.  In a different context, deviations from Porod behaviour are also seen in the microstructure due to mineralization in mammalian cortical bone \cite{mayya2018}.

A well-studied example in the second class is the Ising model in a quenched random field. The degrees of freedom in this case are spins which evolve in time, in a quenched random background. Monte Carlo simulations performed to study ordering dynamics with conserved order parameter reveal that the correlation function exhibits scaling, and shows a cusp singularity at small argument \cite{kumar2017}. 

In a broad sense, the second class of systems has some points of resemblance with a limiting case of fluctuation-dominated order discussed in earlier sections, even though the microscopic dynamics is completely different. In particular, recall the adiabatic limit of particles sliding down fluctuating surfaces, in which particles move on a frozen surface configuration, with the average over configurations being done later.  The earlier described treatments of the CD model (in the case of interacting particles) and random walkers in a frozen Sinai potential (for noninteracting particles) may then be viewed as dealing with problems of quenched disorder.

An interesting point brought out by studies on non-Porod behaviour in disordered systems is that the cusp exponent is related to fractal structure \cite{bale1984, wong1985, gaurav2014}. It would be interesting to establish a similar correspondence in FDPO states, by studying the fractal structure of the disordered regions between long ordered stretches, as in Fig. 3.

\subsection{Other systems which exhibit FDPO and FDC}

Besides the examples of sliding particles on fluctuating surfaces and long-range Ising models that we considered earlier, fluctuation-dominated states of the FDPO and FDC variety arise in diverse situations. Some of these are discussed below.

{\it Passive particles carried by a compressible fluid}.  A highly compressible fluid described by a noisy vortex-free  Burgers equation can be mapped to a surface evolving through KPZ dynamics \cite{medinaKPZ}. This enables results for FDPO and FDC for passive sliders on fluctuating surfaces to be translated, to describe clustering of passive particles in the Burgers fluid.

{\it Lipid clustering in cell membranes}. The collective dynamics of active polar filaments near the surface of a cell can lead to the formation of asters, which form and disappear at a certain rate, but survive long enough to bring about lipid clustering \cite{adas2016}. The mechanism by which lipids are drawn into the center of the aster resembles particles falling into valleys in sliding particle models. A large region in parameter space shows fluctuation-dominated states, with a violation of the Porod law and macroscopic fluctuations in steady state. 

{\it Coarsening in the Vicsek model}. Studies of the coarsening dynamics of the Vicsek model of bird flocking show that there are two distinct values of the dynamic exponent $z$ for the velocity and density fields, and correspondingly, two distinct scaling functions \cite{nisha2020}. Interestingly, the scaling function for density exhibits FDPO with a cusp singularity, while the scaling function for velocity shows a divergence, characteristic of FDC. 

{\it Active nematics} A model of passive particles advected by a nematic field shows FDPO \cite{mishra2006}. For a model of apolar rods whose movement depends on the orientation, Monte Carlo studies suggest that the density of rods shows FDPO \cite{supravat2012}, though a hydrodynamic calculation concludes that active currents can destroy phase separation in two dimensions\cite{shankar2018}.

{\it Vibrated rods}. Experiments on rods vibrated in a tray show giant fluctuations of the density \cite{vnarayan2007}. A subsequent analysis \cite{supravat2012} shows that the data is consistent with standard deviation growing proportionally to the mean, with sub-leading corrections coming from FDC-like corrections in the density.

{\it Inelastically colliding particles}. The coefficient of restitution for inelastically colliding particles depends on the relative velocity of approach. In a model which incorporates this effect, the free collapse of a 1D system of particles results in a state in which the structure functions of density and velocity both depart from the Porod law \cite{shinde2007}. 

\subsection{Conclusion}
FDPO embodies the idea of a phase with long-range order, whose dynamics is dominated by large fluctuations of the order of the mean, resulting in a broad distribution of the order parameter. In most systems which exhibit FDPO, the two-point correlation function is a function of the distance scaled by system size or coarsening length, and the scaling function is singular --- showing a cusp for FDPO, and a divergence for FDC, which describes systems with more intense clustering. These themes --- systems with strong fluctuations, resulting in singularities which may be cusps or divergences --- recall the life and legacy of Michael Fisher.

\section{Acknowledgements}

I gratefully acknowledge collaborations over the years on FDPO and related matters, with Dibyendu Das, Satya Majumdar, Manoj Gopalakrishnan, Sakuntala Chatterjee, Apoorva Nagar, Rajeev Kapri, Malay Bandyopadhyay, Shauri Chakraborti, David Mukamel, Tapas Singha, Samvit Mahapatra, Kabir Ramola, Souvik Sadhukhan, Saroj Nandi and Arghya Das. I am also grateful to Anusheela Chatterjee for useful inputs. This work was supported by the Department of Atomic Energy, India through the DAE Homi Bhabha Chair Professorship and intramural funds at TIFR Hyderabad.



\end{document}